\documentclass{PoS}

\usepackage[utf8]{inputenc}
\usepackage{amsmath}
\usepackage{amsfonts}
\usepackage{amssymb}
\usepackage{bbold}
\usepackage{tikz}
\usepackage{braket}

\usetikzlibrary{positioning,shapes,shadows,arrows,decorations,decorations.pathmorphing,calc}

\newcommand{\ME}{\mathcal{M}}
\newcommand{\Op}{\mathcal{O}}
\newcommand{\unit}{\mathbb{1}}
\newcommand{\dd}{\mathrm{d}}

\title{Two-current correlations and DPDs for the nucleon on the lattice}

\ShortTitle{Two-current correlations and DPDs for the nucleon on the lattice}

\author{\speaker{Christian Zimmermann}\thanks{for RQCD}\\
        Universität Regensburg\\
        E-mail: \email{christian.zimmermann@ur.de}}

\abstract{We calculate correlation functions of two local operators within the nucleon carrying momentum. We resolve their dependence on the spatial distance of the currents. This is carried out for all Wick contractions, taking into account several operator insertion types. The resulting four-point functions can be related to parton distribution functions as well as to Mellin moments of double parton distributions. For the latter, we analyze their quark spin and flavor dependency. In this first study, we employ an $N_F = 2 + 1$ CLS ensemble on a $96 \times 32^3$ lattice with lattice spacing $a = 0.0856\ \mathrm{fm}$ and the pseudoscalar masses $m_\pi = 355\ \mathrm{MeV}$ and $m_K = 441\ \mathrm{MeV}$.}

\FullConference{37th International Symposium on Lattice Field Theory - Lattice2019\\
		16-22 June 2019\\
		Wuhan, China}

\begin{document}

\section{Introduction}

Studying matrix elements of two currents provides the possibility to access parton correlations in a hadron and therefore yields new information about the structure of bound QCD states. These correlations are an important piece of information, e.g. in the context of double parton scattering (DPS) involving double parton distributions (DPDs). Since these effects are significant at energy scales reached at the LHC, the study of two-current matrix elements plays an important role for the determination of the standard model background in the search for new physics. Furthermore, in the short-distance limit, they can be used to directly calculate parton distribution functions (PDFs), see e.g. \cite{Sufian:2019bol}. \\
Two-current matrix elements can be obtained from lattice simulations by calculating four point functions, where the currents are represented by two spatially separated quark bilinears. The corresponding calculations for the pion have been already performed in the past \cite{Bali:2018nde}. \\
In this work we present the current state of our analysis regarding the nucleon (proton). In section \ref{sec:twome} we start with expressing two-current correlations in terms of Euclidean four-point correlators, before we describe the methods for their evaluation in section \ref{sec:sim}. We will provide an overview over all possible Wick contractions and their contributions to physical matrix elements. We continue in section \ref{sec:twist2} with relating Mellin moments of DPDs to Lorentz invariant functions, which we can extract from two-current correlations. We will present some first numerical results before concluding in section \ref{sec:concl}.

\section{Two-current matrix elements on the lattice}
\label{sec:twome}

\begin{figure}
\includegraphics[scale=1]{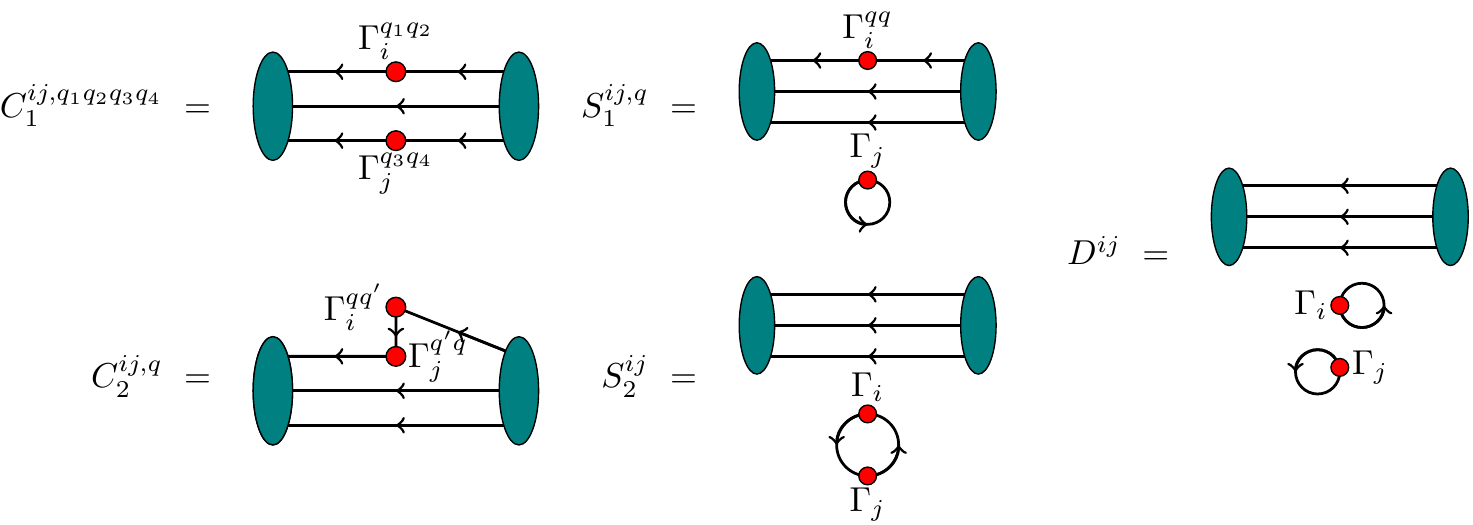}
\caption{Depiction of the five types of Wick contractions (graphs), which contribute to a nucleon two-current matrix element. The explicit Wick contraction depends, in some cases, on the quark flavor of the operator insertions (red points)}
\label{fig:wick}
\end{figure}

Basically, we are interested in matrix elements of a nucleon at momentum $p$ of the following type:
\begin{align}
\ME_{ij}(p,y) := \bra{p} \Op_j(y)\ \Op_i(0) \ket{p}\ .
\label{def:TwoCME}
\end{align}
The two operator insertions separated by a vector $y$ are considered to be local quark bilinears $\bar{q} \Gamma_i q$ of definite flavor $q$. For $\Gamma_i$ we take the following Dirac structures: $S=\unit$ (scalar), $P=i\gamma_5$ (pseudoscalar), $V^\mu=\gamma^\mu$ (vector), $A^\mu=\gamma^\mu \gamma_5$ (axial vector) and $T^{\mu\nu}=\sigma^{\mu\nu}$ (tensor). At the moment we do not consider the hadron's polarization and always take the spin average of the matrix elements. \\
We are able to evaluate the matrix elements \eqref{def:TwoCME} on an Euclidean lattice under the condition that there is no separation in time direction, $y^0 = 0$. An explicit formula is given by:
\begin{align}
\ME_{ij}(p,y)|_{y^0 = 0} = 2V_3 \sqrt{m^2 + \vec{p}^2} \left. \frac{C_{\mathrm{4pt}}(\vec{y},t,\tau)}{C_\mathrm{2pt}(t)} \right|_{0 \ll \tau \ll t}\ ,
\label{rel:ME_latt}
\end{align}
where the four point function $C_{\mathrm{4pt}}$ and the two point function $C_{\mathrm{2pt}}$ are defined as:
\begin{align}
C_{\mathrm{4pt}}(\vec{y},t,\tau) = \langle \mathcal{P}^p(t)\ \Op_j(\vec{y},\tau)\ \Op_i(\vec{0},\tau)\ \overline{\mathcal{P}}^p(0) \rangle\ , \qquad
C_{\mathrm{2pt}}(t) = \langle \mathcal{P}^p(t)\ \overline{\mathcal{P}}^p(0) \rangle\ .
\end{align}
It is understood that the nucleon interpolators $\overline{\mathcal{P}}^p$, $\mathcal{P}^p$, projecting onto momentum $p$ and positive parity, also overlap with excited states, which are aimed to be suppressed by the limit given in \eqref{rel:ME_latt}. 
%Explicitly we will use for the interpolators:
%\begin{align}
%\mathcal{P}^{\vec{p}}(t) = \left.\frac{1}{2} \sum_{\vec{x}} e^{-i\vec{x}\cdot\vec{p}} \epsilon_{abc} (1+\gamma_4) u_a(x) [ u_b^T(x) i \gamma_2 \gamma_4 \gamma_5 d_c(x) ] \right|_{t=x_4} \\
%\overline{\mathcal{P}}^{\vec{p}}(t) = \left.\frac{1}{2} \sum_{\vec{x}} e^{i\vec{x}\cdot\vec{p}} \epsilon_{abc} [ \bar{u}_a^T(x) i \gamma_2 \gamma_4 \gamma_5 \bar{d}_b(x) ] \bar{u}_c(x) (1+\gamma_4)   \right|_{t=x_4}
%\end{align}
For the evaluation of $C_{4\mathrm{pt}}$ we have to consider five types of Wick contractions. Following the nomenclature of \cite{Bali:2018nde} we call the contractions of a certain topology $C_1$, $C_2$, $S_1$, $S_2$ and $D$, see Figure \ref{fig:wick}. Depending on the explicit quark flavor a definite set of contractions contributes to each matrix element. Taking into account only light quark operator insertions, we can formulate the following decompositions for proton-proton matrix elements:
\begin{align}
\begin{aligned}
\bra{p} \Op_j^{dd}(\vec{y})\ \Op_i^{uu}(\vec{0}) \ket{p} &= C^{ij,uudd}_1(\vec{y}) + S^{ij,u}_1(\vec{y}) + S^{ji,d}_1(-\vec{y}) + D^{ij}(\vec{y})\ , \\
\bra{p} \Op_j^{uu}(\vec{y})\ \Op_i^{uu}(\vec{0}) \ket{p} &= C^{ij,uuuu}_1(\vec{y}) + C^{ij,u}_2(\vec{y}) + C^{ji,u}_2(-\vec{y}) + S^{ij,u}_1(\vec{y}) + S^{ji,u}_1(-\vec{y}) \\
 &+ S_2^{ij}(\vec{y}) + D^{ij}(\vec{y})\ , \\
 \bra{p} \Op_j^{dd}(\vec{y})\ \Op_i^{dd}(\vec{0}) \ket{p} &= C^{ij,d}_2(\vec{y}) + C^{ji,d}_2(-\vec{y}) + S^{ij,d}_1(\vec{y}) + S^{ji,d}_1(-\vec{y}) + S_2^{ij}(\vec{y}) + D^{ij}(\vec{y})\ , \\
 \bra{p} \Op_j^{du}(\vec{y})\ \Op_i^{ud}(\vec{0}) \ket{p} &= C^{ij,uddu}_1(\vec{y}) + C^{ij,u}_2(\vec{y}) + C^{ji,d}_2(-\vec{y}) + S_2^{ij}(\vec{y})\ .
\end{aligned}
\label{rel:phys_me}
\end{align}
In this work we will focus on the three matrix elements, which contain only flavor conserving operators.

\section{Simulation details and results}
\label{sec:sim}

\begin{table}
\begin{center}
\begin{tabular}{llllllllll}
\hline
\hline
id & $\beta$ & $a[\mathrm{fm}]$  & $L^3 \times T$ & $\kappa_{l/s}$ & $m_{\pi / K }[\mathrm{MeV}]$ & $m_\pi L$ & conf. \\
\hline
H102 & $3.4$ & $0.0856$ & $32^3 \times 96$ & $0.136865$ & $355$ & $4.9$ & $2037$ \\
 & & & & $0.136549339$ & $441$ & & \\
\hline
\hline
\end{tabular}
\end{center}
\caption{Details of the CLS ensemble used for our simulation. At the moment, the simulation only includes 960 of the configurations.\label{tab:cls}}
\end{table}

\begin{table}
\begin{center}
\begin{tabular}{l|llll}
\hline
\hline
 & $C_1$ & $C_2$ & $S_1$(loop) & $D$ (2 loops) \\ 
\hline
$N_\mathrm{stoch}$ & $2$ & $96$ & $120$ & $60$ \\
\hline
\hline
\end{tabular}
\end{center}
\caption{Number of stochastic noise vectors used for the inversion of the stochastic propagators for each graph.\label{tab:stochnum}}
\end{table}

\begin{figure}
\begin{minipage}[t]{8cm}
(a)\\
\includegraphics[scale=0.205]{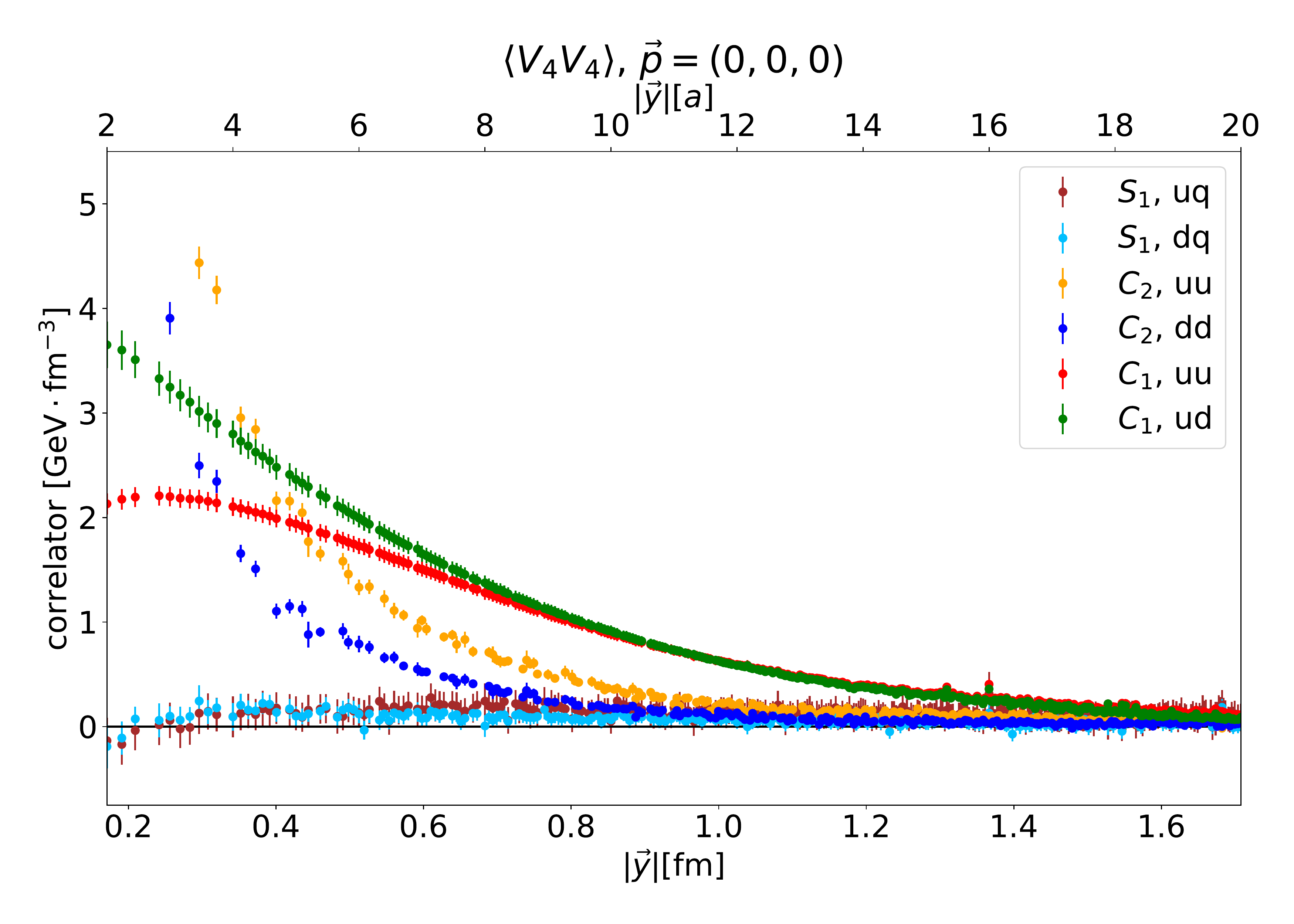}
\end{minipage}
\begin{minipage}[t]{8cm}
(b)\\
\includegraphics[scale=0.205]{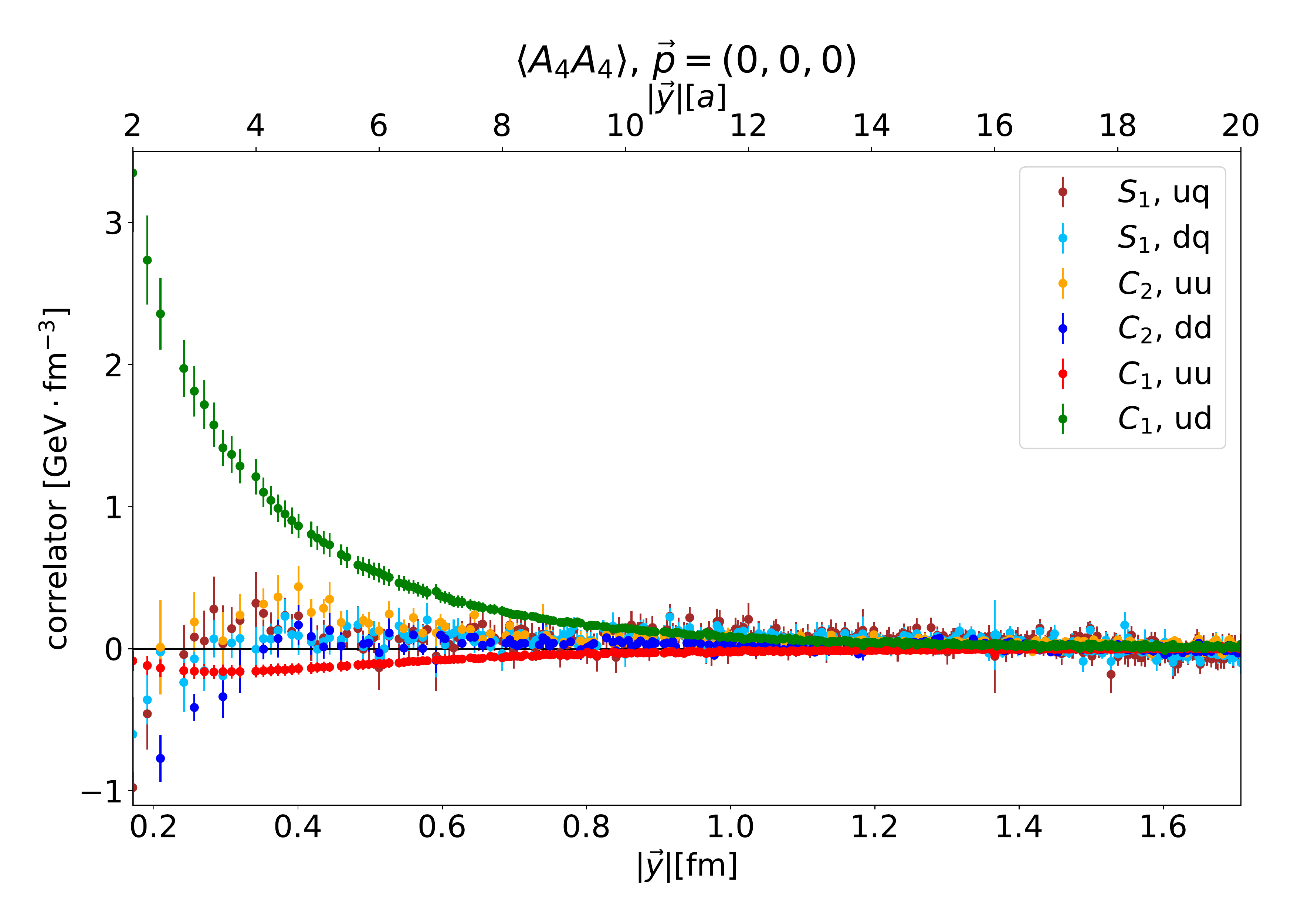}
\end{minipage}
\caption{Results for two-current matrix elements for operator insertions $V^4 V^4$ (a) and $A^4 A^4$ (b) in dependence of the operator distance $|\vec{y}|$. The data are shown for the graphs $C_1$, $C_2$ and $S_1$ for a specific flavor combination. $q$ can be both, an up or a down quark. \label{res:me_graph}}
\end{figure}

We perform our simulation on a CLS ensemble employing $n_f = 2+1$ $\Op(a)$-improved Wilson fermions \cite{Bruno:2014jqa}, where we use 960 configurations. The details of the ensemble are listed in Table \ref{tab:cls}. To increase the overlap of the interpolators with the nucleon ground state at specific momentum, we employ the Wuppertal smearing technique on boosted sources, also known as momentum smearing \cite{Bali:2016lva}, where we use APE-smeared gauge links \cite{Falcioni:1984ei}. \\
We start our analysis by measuring the graphs $C_1$, $C_2$, $S_1$ and $D$. The disconnected contribution $S_2$ will be treated in the near future. All graphs are evaluated on a nucleon source located at a random spatial position. Since the used ensemble has open boundary conditions in the time direction, we choose the source time slice to be at $t_{\mathrm{src}} = T/2 = 48a$. For $C_1$, $C_2$, $S_1$ we have at least one propagator connecting the nucleon sink with an insertion operator. This propagator we get from an inversion on a sequential source \cite{Martinelli:1988rr}. In the case of $C_1$ this sequential source contains also the source of a stochastic propagator, which is needed to connect the sink with the second insertion. The additional propagator connecting the two insertions in $C_2$ is obtained by inversion on a stochastic source. Furthermore it is improved by removing $n = |y_1|+|y_2|+|y_3|$ hopping terms, which do not contribute, but, nevertheless, may create noise \cite{Bali:2009hu}. The same is done for the stochastic propagators of the loops in $S_1$ and $D$, where the number of hopping terms, which can be neglected, depends on the specific insertion type. An overview of the number of stochastic noise vectors we used for each propagator can be found in Table \ref{tab:stochnum}.\\
The structure of $C_1$ allows calculations for many insertion time slices simultaneously, while we have to fix the insertion time for all other graphs. Otherwise, we would have to repeat the calculation for each insertion time. Keeping the source-sink-separation fixed at $t = t_{\mathrm{snk}} - t_{\mathrm{src}} = 12a$ for zero momentum and $t=10a$ otherwise, we use as insertion time $\tau = t_{\mathrm{ins}} - t_{\mathrm{src}} = t/2$ for the graphs $C_2$, $S_1$ and $D$. For $C_1$ we perform a fit over the insertion time, where we take into account the range $3a \le \tau \le t-3a$. As nucleon momenta we choose $\vec{p} = \vec{k}\cdot 2\pi/L$, with $\vec{k}\in \{(0,0,0),(1,1,1),(2,2,2)\}$. We renormalize our results and convert them to the $\overline{\mathrm{MS}}$-scheme, see e.g. \cite{Bali:2019ecy}.\\
Basically we take into account the insertion combinations $SS$, $PP$, $V^\mu V^\nu$, $A^\mu A^\nu$, $V^\mu T^{\nu \rho}$ and $T^{\mu\nu} T^{\rho\sigma}$. As a first look into the results we focus on the matrix elements for $V^4 V^4$ and $A^4 A^4$, which are rotationally invariant. The data as functions of the spatial operator insertion distance $|\vec{y}|$ for the graphs $C_1$, $C_2$ and $S_1$ are shown in Figure \ref{res:me_graph}. We do not show the result for the $D$ contraction, since it does not yield a clear signal. For the vector channel we observe a clear signal for both connected contractions, where their contributions differ w.r.t. the quark flavor. Comparing $C_1$ and $C_2$, we can state that $C_2$ is mainly relevant for small distances, while it decreases very fast for large distance. The signal of $S_1$ is comparable with zero. The axial vector channel is dominated by the $ud$ contribution of the $C_1$ contraction.

\section{Twist-2 operator analysis and DPDs}
\label{sec:twist2}

\begin{figure}
\begin{minipage}[t]{8cm}
(a)\\
\includegraphics[scale=0.21]{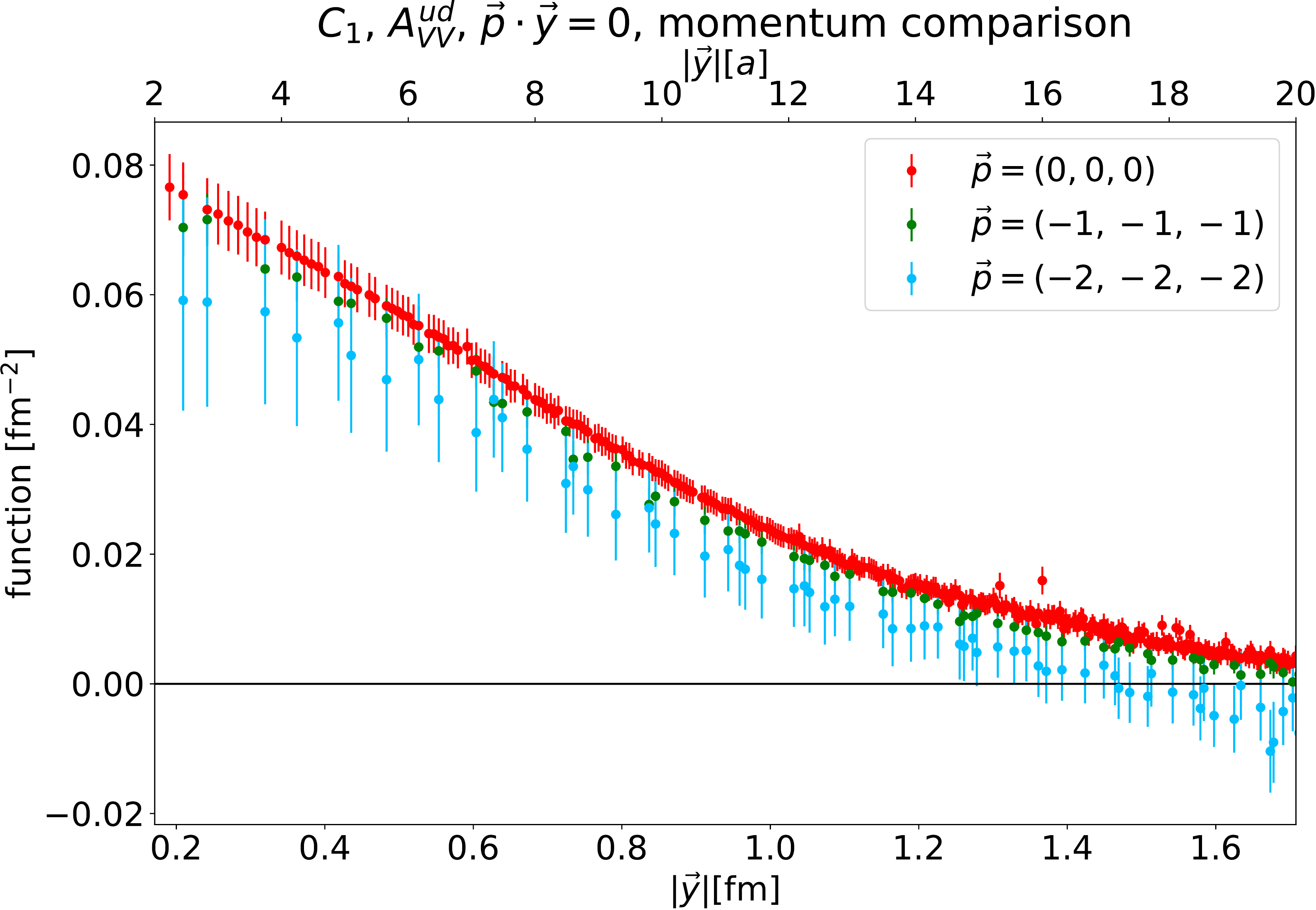}
\end{minipage}
\begin{minipage}[t]{8cm}
(b)\\
\includegraphics[scale=0.21]{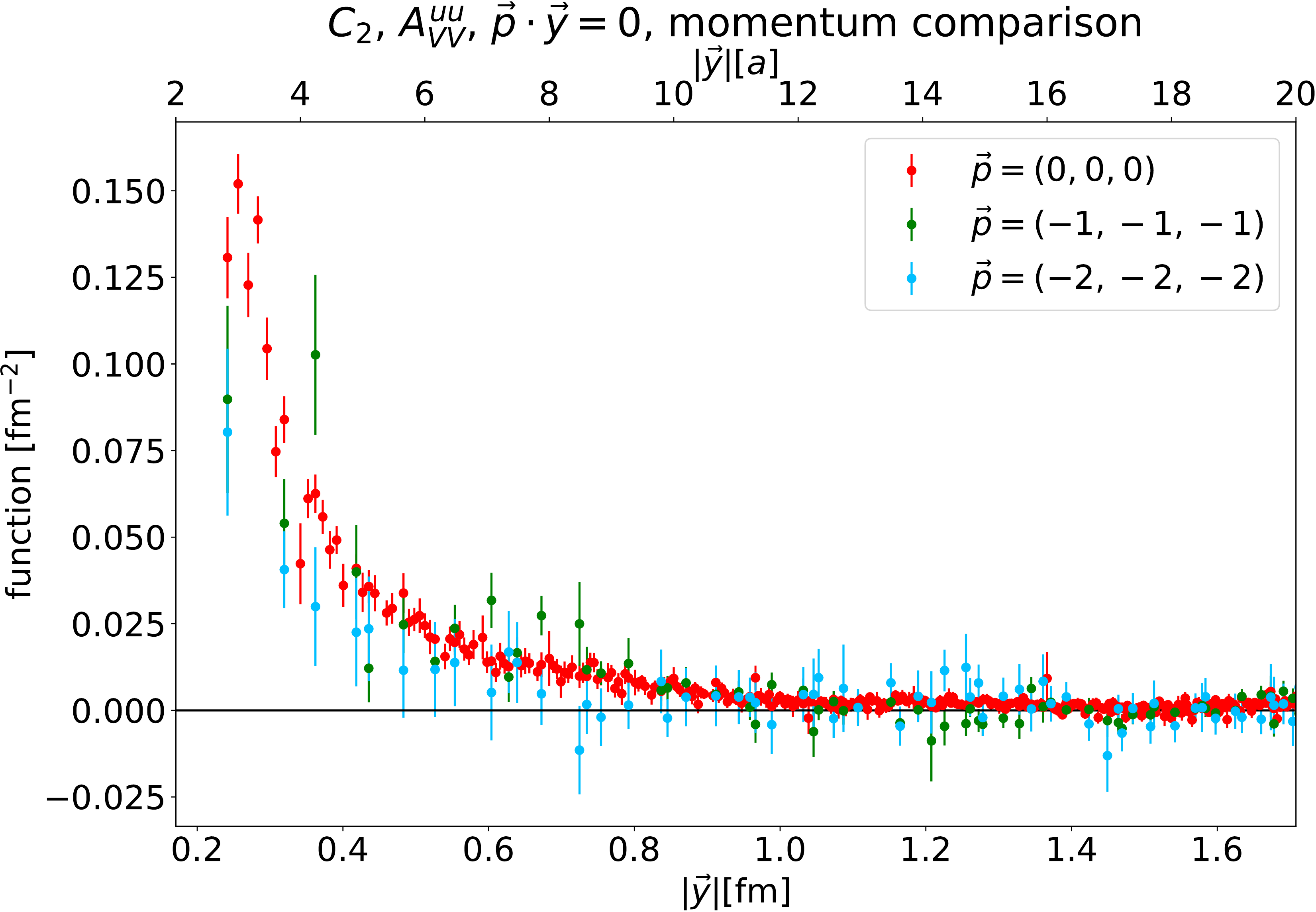}
\end{minipage}
\caption{Comparison between different momenta for the invariant function $A(p\cdot y=0,y^2)$. The data are shown for the graph $C_1$ and quark flavor $ud$ (a) and the graph $C_2$ with flavor $uu$ (b). In both cases the data appear to be consistent with Lorentz invariance. \label{res:inv_mom}}
\end{figure}

\begin{figure}
\begin{minipage}[t]{8cm}
(a)\\
\includegraphics[scale=0.21]{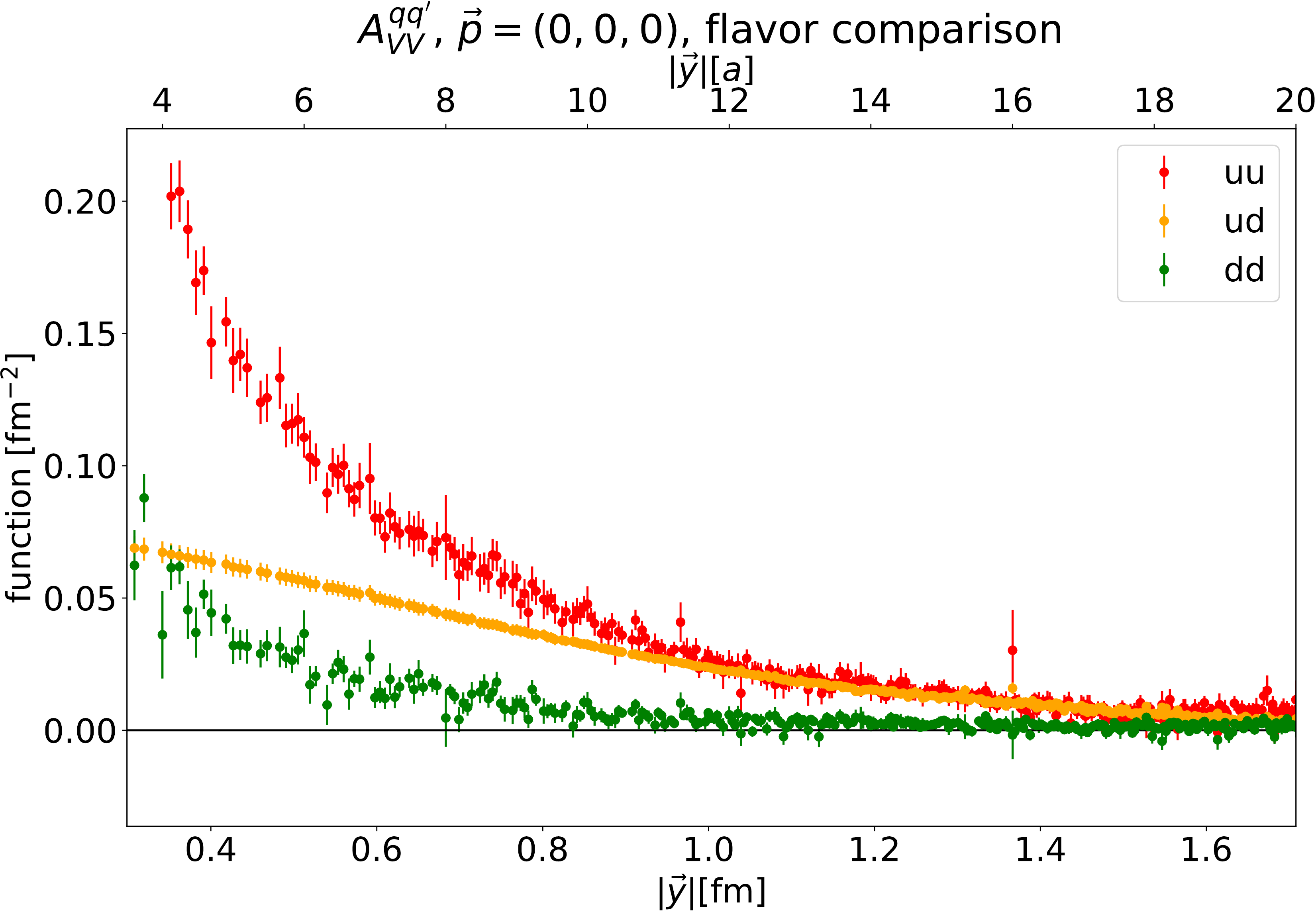}
\end{minipage}
\begin{minipage}[t]{8cm}
(b)\\
\includegraphics[scale=0.21]{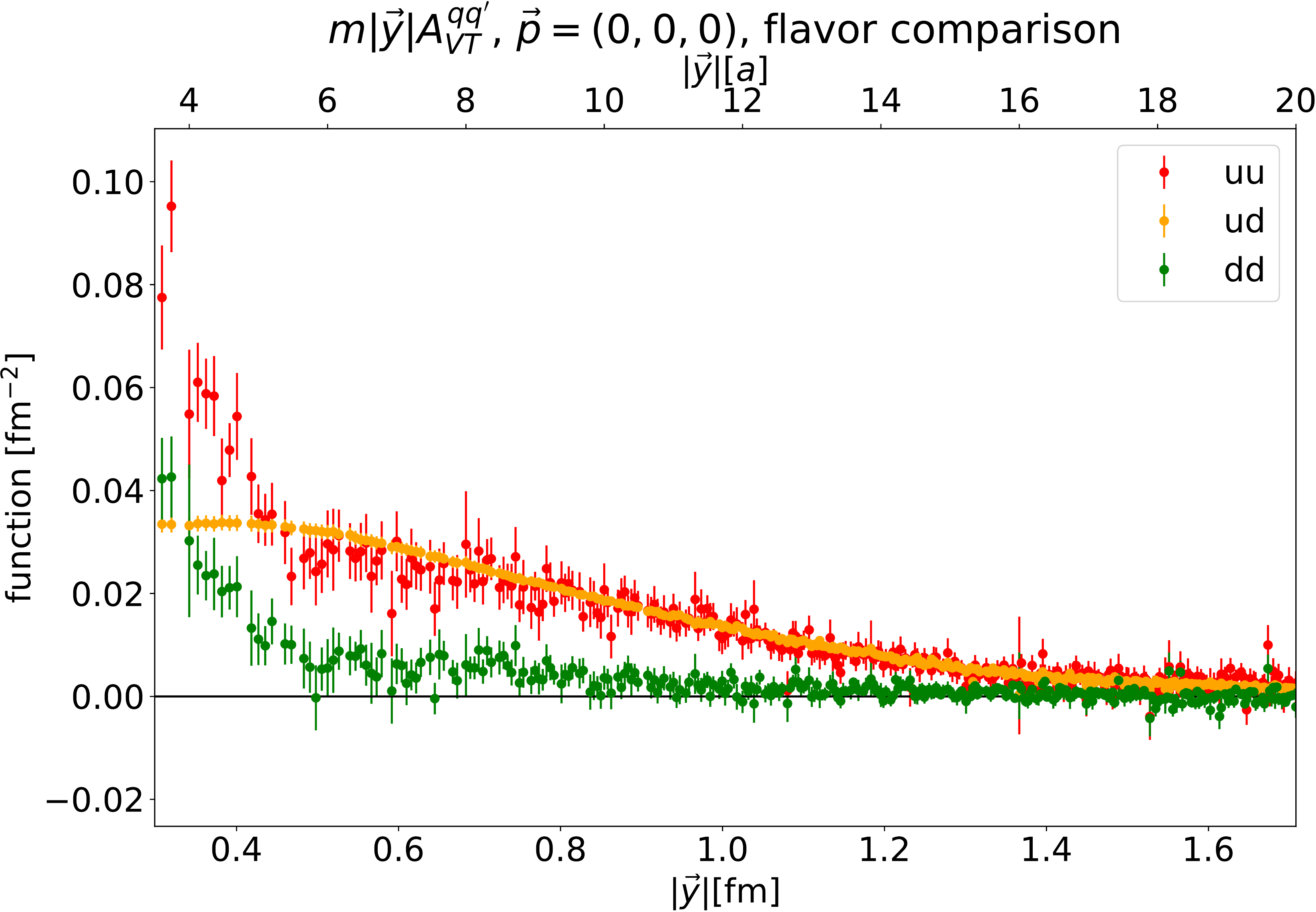}
\end{minipage}
\caption{Physical combination of graphs for $A(p\cdot y=0,y^2)$, where the signal of different flavor combinations is compared for the channels $VV$ (a) and $VT$ (b). Disconnected contributions are neglected. \label{res:inv_flav}}
\end{figure}

\begin{figure}
\begin{minipage}[t]{8cm}
(a)\\
\includegraphics[scale=0.21]{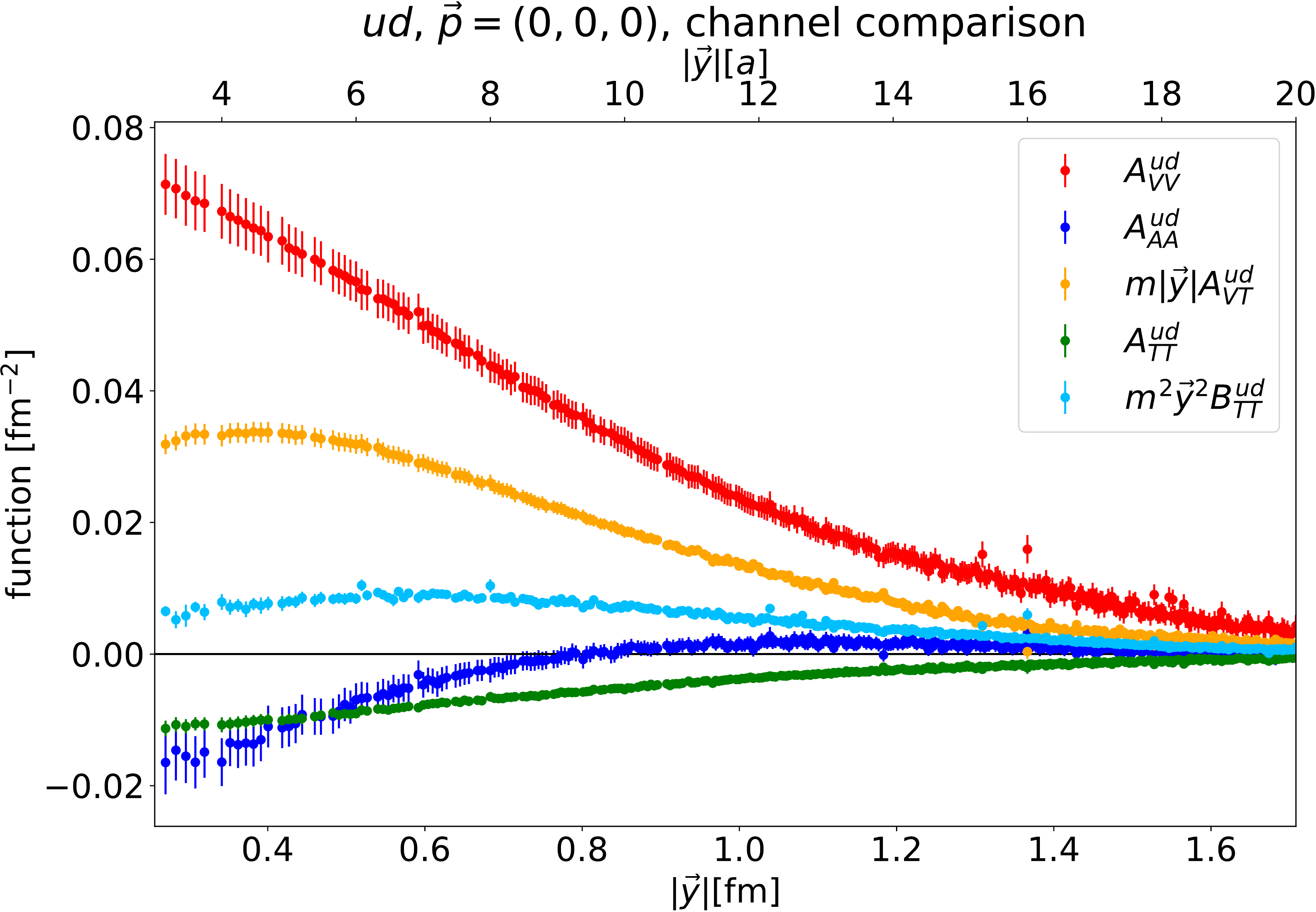}
\end{minipage}
\begin{minipage}[t]{8cm}
(b)\\
\includegraphics[scale=0.21]{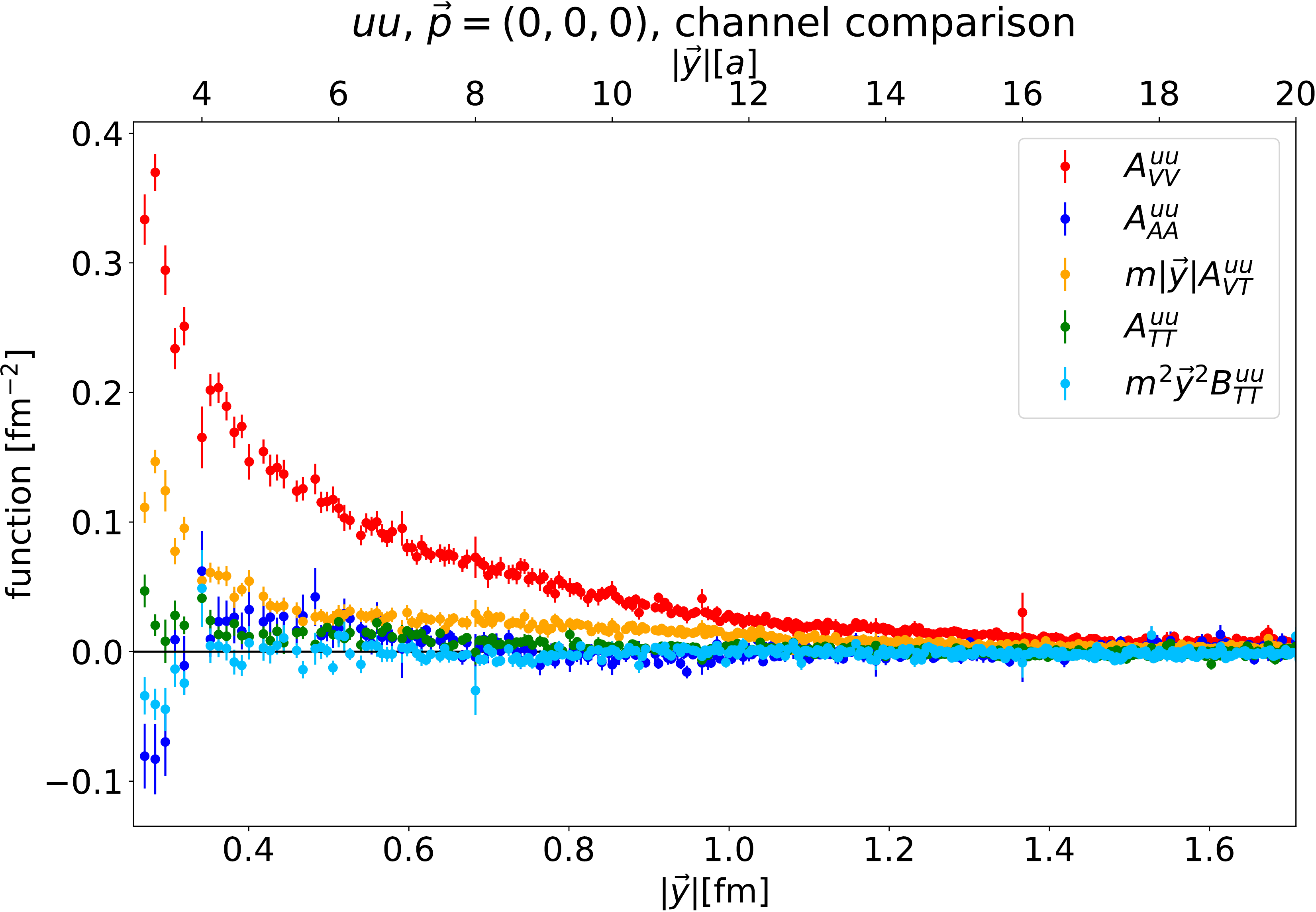}
\end{minipage}
\caption{Comparison between different insertion types corresponding to a certain quark polarization for flavor $ud$ (a) and flavor $uu$ (b). Disconnected contributions are neglected. \label{res:inv_chan}}
\end{figure}

In the following we explain the analysis regarding twist-2 operators. In the context of the parton model, the corresponding matrix elements are related to the probability of finding two quarks of a certain polarization. Explicitly, we consider operators of the type $\bar{q} \Gamma q$, where $\Gamma$ is $\gamma^+$ (unpolarized quarks), $\gamma^+ \gamma_5$ (longitudinally polarized) or $\sigma^{+j}$ (transversely polarized). We use the property that our matrix elements are Lorentz tensors, which depend only on $p$ and $y$, to project them onto a couple of Lorentz invariant functions. This additionally increases the statistics. 
%Furthermore, we subtract the trace to reduce the degrees of freedom, which is denoted by $\mathcal{T}$.
%\begin{align}
%\begin{aligned}
%\mathcal{T} \mathcal{M}_{VV/AA}^{\mu\nu}(p,y) &= \left(2p^\mu p^\nu - \frac{p^2}{2}g^{\mu\nu}\right) A(p\cdot y,y^2) + \left(2p^{\{\mu} y^{\nu\}} - \frac{p\cdot y}{2}g^{\mu\nu}\right) m^2 B(p\cdot y,y^2) \\
%&+ \left(2y^\mu y^\nu - \frac{y^2}{2}g^{\mu\nu}\right) m^4 C(p\cdot y,y^2) 
%\end{aligned}
%\end{align}
For leading twist matrix elements we see that only one invariant function $A(p\cdot y,y^2)$ contributes for $VV$, $AA$ and $VT$, whereas in the case of two tensor currents there is also the quadrupole function $B(p\cdot y,y^2)$ contributing:
\begin{align}
\begin{aligned}
\mathcal{M}_{VV/AA}^{++}(p,y) &= 2(p^+)^2 A_{VV/AA}(p\cdot y,y^2)\ , \qquad \mathcal{M}_{VT}^{+j+}(p,y) = 2(p^+)^2 m\ y^j A_{VT}(p\cdot y,y^2)\ , \\
\mathcal{M}_{TT}^{j+l+}(p,y) &= 2(p^+)^2 \left[ \delta^{jl} A_{TT}(p\cdot y,y^2) - m^2 \left( \delta^{jl}y^2 + 2y^j y^l \right) B_{TT}(p\cdot y,y^2) \right]\ .
\end{aligned}
\end{align}
The fact that we simulate on an Euclidean lattice restricts the accessibility of $A(p\cdot y,y^2)$ and $B(p\cdot y,y^2)$ to the region, where $-y^2 = \vec{y}^2 \ge 0$ and $(p\cdot y)^2 = (\vec{p}\cdot\vec{y})^2 \le \vec{p}^2 \vec{y}^2$.\\
After integrating over $y^-$, these functions can be related to the first Mellin moment in each momentum fraction $I(\vec{y}_\perp^2)$ of DPDs $f(x_i,\vec{y}_\perp^2)$, see section 4 of \cite{Diehl:2011yj}\footnote{Comparing to our notation regarding the quark polarization, there is the correspondence $q \leftrightarrow V$, $\Delta q \leftrightarrow A$, $\delta q \leftrightarrow T$}. For $\vec{p}_\perp = \vec{0}$ we can integrate over $p\cdot y$:
\begin{align}
%\begin{aligned}
I_{ij}(\vec{y}_\perp^2) = \int \dd (p\cdot y)\ A_{ij}(p\cdot y,y^2)\ , \qquad
I_{TT}^t(\vec{y}_\perp^2) = \int \dd (p\cdot y)\ B_{TT}(p\cdot y,y^2)\ .
%\end{aligned}
\end{align}
This extraction is left for future work. In the following we will only present results for the relevant invariant functions. As described in the last section our calculation was carried out for several momenta. Some of the results for $p\cdot y=0$ are shown in Figures \ref{res:inv_mom}, \ref{res:inv_flav}, \ref{res:inv_chan}. In Figure \ref{res:inv_mom} we compare the data for different momenta, first for graph $C_1$ and quark flavor $ud$ (a), second for graph $C_2$ with flavor $uu$ (b). Here we observe good consistency with Lorentz invariance. Since the data for $\vec{p} \neq \vec{0}$ are mainly relevant for the actual extraction of DPDs, which is not discussed in this work, we will concentrate on zero momentum in the following. Figure \ref{res:inv_flav} shows the dependency on the quark flavor for the $VV$ (a) and $VT$ (b) case. Here we already take physical sums of graphs according to \eqref{rel:phys_me}, but neglect disconnected contributions, such as $S_1$ or $D$. In both cases one finds that the $uu$ and $ud$ signals dominate for larger distances. For small $\vec{y}$ the signal of $uu$ and $dd$ is strongly increasing.\\
Let us now take a closer look at the channel dependency of our data, i.e. the effects of quark polarization. The corresponding data are plotted in Figure \ref{res:inv_chan} for quark flavor $uu$ and $ud$. The contribution of unpolarized quarks is dominant in both cases. For the flavor combination $ud$ one can observe a non-negligible signal for all other polarization types, whereas in the $uu$ case these effects appear to be suppressed.

\section{Conclusion}
\label{sec:concl}

We considered two-current matrix elements for the proton on the lattice, where we evaluated four out of five contributing Wick contractions, resolving the dependency on the quark distance. This has been done for several flavor combinations and insertion types. For all but the $D$ graph, which is very noisy, the statistical precision of the data is very promising. Furthermore, we extracted Lorentz invariant functions, which are related to Mellin moments of DPDs, where we compared several flavor combinations as well as operator insertion types corresponding to a certain quark polarization. For the latter we found dominance of unpolarized quarks, but non-negligible polarization effects for the flavor combination $ud$.\\
It remains to evaluate one missing Wick contraction ($S_2$). Furthermore, we aim to analyze the data for non-zero $p\cdot y$ and finally obtain results for DPD Mellin moments. It is also planned to repeat the analysis on larger lattices closer to the physical point.

\section*{Acknowledgments}

I thank the RQCD collaboration and especially A. Schäfer, M. Diehl and G. S. Bali for fruitful discussions, which were very helpful for the calculations and simulations. Furthermore, I thank B. Gläßle for technical support. I also acknowledge the CLS effort of generating the $N_f = 2 + 1$ ensembles. The simulations have been performed on the SFB/TRR 55 QPACE 3.

\bibliographystyle{bst/jbJHEP.bst}

\begin{thebibliography}{99}

\bibitem{Sufian:2019bol}
  R.~S.~Sufian, J.~Karpie, C.~Egerer, K.~Orginos, J.~W.~Qiu and D.~G.~Richards,
  %``Pion Valence Quark Distribution from Matrix Element Calculated in Lattice QCD,''
  Phys.\ Rev.\ D {\bf 99} (2019) no.7,  074507
  %doi:10.1103/PhysRevD.99.074507
  [arXiv:1901.03921 [hep-lat]].
  %%CITATION = doi:10.1103/PhysRevD.99.074507;%%
  %14 citations counted in INSPIRE as of 24 Sep 2019
  
\bibitem{Bali:2018nde}
  G.~S.~Bali {\it et al.},
  %``Two-current correlations in the pion on the lattice,''
  JHEP {\bf 1812} (2018) 061
  %doi:10.1007/JHEP12(2018)061
  [arXiv:1807.03073 [hep-lat]].
  %%CITATION = doi:10.1007/JHEP12(2018)061;%%
  %6 citations counted in INSPIRE as of 24 Sep 2019

\bibitem{Bruno:2014jqa}
  M.~Bruno {\it et al.},
  %``Simulation of QCD with N$_{f} =$ 2 $+$ 1 flavors of non-perturbatively improved Wilson fermions,''
  JHEP {\bf 1502} (2015) 043
  %doi:10.1007/JHEP02(2015)043
  [arXiv:1411.3982 [hep-lat]].
  %%CITATION = doi:10.1007/JHEP02(2015)043;%%
  %134 citations counted in INSPIRE as of 25 Sep 2019

\bibitem{Bali:2016lva}
  G.~S.~Bali, B.~Lang, B.~U.~Musch and A.~Schäfer,
  %``Novel quark smearing for hadrons with high momenta in lattice QCD,''
  Phys.\ Rev.\ D {\bf 93} (2016) no.9,  094515
  %doi:10.1103/PhysRevD.93.094515
  [arXiv:1602.05525 [hep-lat]].
  %%CITATION = doi:10.1103/PhysRevD.93.094515;%%
  %74 citations counted in INSPIRE as of 25 Sep 2019

\bibitem{Falcioni:1984ei}
  M.~Falcioni, M.~L.~Paciello, G.~Parisi and B.~Taglienti,
  %``Again On Su(3) Glueball Mass,''
  Nucl.\ Phys.\ B {\bf 251} (1985) 624.
  %doi:10.1016/0550-3213(85)90280-9
  %%CITATION = doi:10.1016/0550-3213(85)90280-9;%%
  %170 citations counted in INSPIRE as of 30 Sep 2019
  
\bibitem{Martinelli:1988rr}
  G.~Martinelli and C.~T.~Sachrajda,
  %``A Lattice Study of Nucleon Structure,''
  Nucl.\ Phys.\ B {\bf 316} (1989) 355.
  %doi:10.1016/0550-3213(89)90035-7
  %%CITATION = doi:10.1016/0550-3213(89)90035-7;%%
  %132 citations counted in INSPIRE as of 30 Sep 2019
  
\bibitem{Bali:2009hu}
  G.~S.~Bali, S.~Collins and A.~Schäfer,
  %``Effective noise reduction techniques for disconnected loops in Lattice QCD,''
  Comput.\ Phys.\ Commun.\  {\bf 181} (2010) 1570
  %doi:10.1016/j.cpc.2010.05.008
  [arXiv:0910.3970 [hep-lat]].
  %%CITATION = doi:10.1016/j.cpc.2010.05.008;%%
  %134 citations counted in INSPIRE as of 25 Sep 2019  
  
\bibitem{Bali:2019ecy}
  G.~S.~Bali {\it et al.} [RQCD Collaboration],
  %``Light-cone distribution amplitudes of octet baryons from lattice QCD,''
  Eur.\ Phys.\ J.\ A {\bf 55} (2019) no.7,  116
  %doi:10.1140/epja/i2019-12803-6
  [arXiv:1903.12590 [hep-lat]].
  %%CITATION = doi:10.1140/epja/i2019-12803-6;%%
  %3 citations counted in INSPIRE as of 30 Sep 2019

\bibitem{Diehl:2011yj}
  M.~Diehl, D.~Ostermeier and A.~Schäfer,
  %``Elements of a theory for multiparton interactions in QCD,''
  JHEP {\bf 1203} (2012) 089
   Erratum: [JHEP {\bf 1603} (2016) 001]
  %doi:10.1007/JHEP03(2012)089, 10.1007/JHEP03(2016)001
  [arXiv:1111.0910 [hep-ph]].
  %%CITATION = doi:10.1007/JHEP03(2012)089, 10.1007/JHEP03(2016)001;%%
  %210 citations counted in INSPIRE as of 25 Sep 2019
  
%\bibitem{Goeckeler:priv}
%  M.~Göckeler,
%  \textit{private communication}



\end{thebibliography}

\end{document}